\documentclass[aps,preprint,prd,showpacs,nofootinbib]{revtex4}

\usepackage{latexsym}
\usepackage{amsmath}
\usepackage{graphicx}
\usepackage{subfigure}
\usepackage{dcolumn}
\usepackage{bm}
\usepackage{amssymb}
\usepackage{latexsym}
\usepackage{ulem}
\usepackage{color}
\usepackage[colorlinks,linkcolor=magenta,anchorcolor=cyan,citecolor=blue]{hyperref}

\def\be{\begin{equation}}
\def\ee{\end{equation}}
\def\ba{\begin{eqnarray}}
\def\ea{\end{eqnarray}}

\def\nn{\nonumber}
\def\lf{\left}
\def\rt{\right}

\begin{document}

\title{ Are gravitational wave ringdown echoes always equal-interval?}

\author{Yu-Tong Wang$^{1}$\footnote{wangyutong@ucas.ac.cn}}
\author{Zhi-Peng Li$^{1}$\footnote{lizhipeng172@mails.ucas.ac.cn}}
\author{Jun Zhang$^{2}$$\footnote{jun34@yorku.ca}$}
\author{Shuang-Yong Zhou$^{3}$\footnote{zhoushy@ustc.edu.cn}}
\author{Yun-Song Piao$^{1,4}$\footnote{yspiao@ucas.ac.cn}}

\affiliation{$^1$ School of Physics, University of Chinese Academy
of Sciences, Beijing 100049, China}

\affiliation{$^2$Department of Physics and Astronomy, York University, Toronto, Ontario, M3J 1P3, Canada}

\affiliation{$^3$  Interdisciplinary Center for Theoretical Study,
University of Science and Technology of China, Hefei, Anhui 230026, China}

\affiliation{$^4$ Institute of Theoretical Physics, Chinese
Academy of Sciences, P.O. Box 2735, Beijing 100190, China}

\begin{abstract}

Gravitational wave (GW) ringdown waveforms may
contain ``echoes'' that encode new physics in the
strong gravity regime. It is commonly assumed that
the new physics gives rise to the GW echoes whose intervals are
constant. We point out that this assumption is not
always applicable.
In particular, if the post-merger object is initially a
wormhole, which slowly pinches off and eventually collapses
into a black hole, the late-time ringdown waveform exhibit
a series of echoes whose intervals are increasing
with time. We also assess how this affects the
ability of Advanced LIGO/Virgo to detect these new
signals.

\end{abstract}

\maketitle

\section{Introduction}

Recently, the LIGO Scientific and Virgo Collaborations, using
 ground based laser interferometers, have
detected gravitational wave (GW) signals of binary black hole (BH)
\cite{Abbott:2016blz} and binary neutron stars
\cite{TheLIGOScientific:2017qsa} coalescences, which opened
a new window to probe gravity physics, particularly in the strong
field regime, and the origin of universe.

Inflation is the current paradigm of the early universe.
The domain-wall bubbles (or relevant objects) can
spontaneously nucleate in de Sitter space and be stretched by the
inflation to astrophysical scales \cite{Basu:1991ig}, see
also \cite{Zhang:2015bga}. In Refs.\cite{Garriga:2015fdk,
Deng:2016vzb}, it has been argued that under certain conditions
the interior of a large bubble will develop into a
baby universe, which is connected to the exterior region through a
wormhole (WH), see also \cite{Kodama:1981gu}. The throat of
the WH is dynamic, which will pinch off shortly after the WH
 enters into the cosmological horizon, or see
\cite{Roman:1992xj}. The resulting BHs might be
candidates for seeding the supermassive objects at the
center of galaxies \cite{Kormendy:1995er}. Thus, it is
possible that the primordial WHs, created and enlarged in the
inflationary phase, might be slowly pinching till today,
and merge with another compact object (a
neutron star or BH). In any case, one
could speculate a scenario where a WH may appear as an
intermediate state in the coalescences of some compact
objects (BH/BH, WH/BH, WH/WH mergers, etc.).

In this paper, we will show that if the post-merger object
is a WH, which is slowly pinching off (and eventually will
collapse into a black hole), the late-time ringdown waveform will
exhibit a series of interval-increasing echoes. It
is commonly assumed, after Cardoso et.al.'s seminal
work Refs.\cite{Cardoso:2016rao, Cardoso:2016oxy}, that the
intervals between the neighboring GW echoes are constant,
which has been widely used in searching for the signals of echoes
in GW data\cite{Abedi:2016hgu, Abedi:2017isz,
Westerweck:2017hus}. However, this assumption could
bring bias that causes systematic errors in the parameter
estimation of signals, e.g.\cite{Maselli:2017tfq},
as we find that the GW echoes may not
be equal-interval.
Our result suggests a more general pool of
templates for the echo searches might be desirable.



\section{Setup and Ringdown Echoes}

\begin{figure}[htbp]
\includegraphics[width=.48\textwidth]{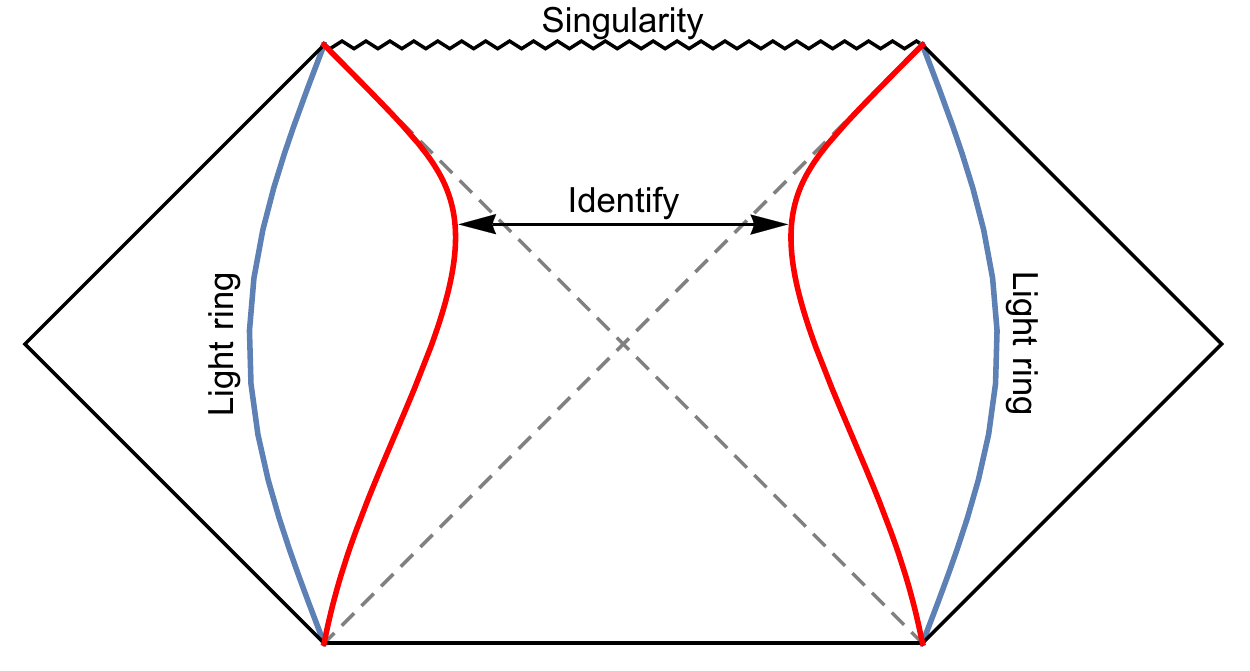}
\caption{The conformal diagram of a slowly pinching WH. The WH is
constructed by gluing two Schwarzschild space-time at $r=r_0(t)$
(the red lines), which start at somewhere inside the light ring
and end at the Schwarzschild radius at a finite $t$. The blue
lines show the light rings of the Schwarzschild metrics.}
\label{fig02}
\end{figure}

Let us begin with the spacetime depicted by
Fig.\ref{fig02}, where the post-merger object is
initially a WH, which slowly pinches for a period before
collapsing into a BH. Here, for our purposes,  we
make use of the simple phenomenological model, the Morris-Thorne
WH \cite{Morris:1988cz, Morris:1988tu}, which is obtained by
gluing the Schwarzschild metrics
\begin{align}
ds^{2}=-B dt^{2}+{dr^{2}\over B}+r^{2}d\Omega^{2},\quad
\lf(B=1-{2M\over r}\rt)
\end{align}
of both sides at $r=r_{0}>2M$, where $r_{0}$ is the radius of the
throat, see e.g.\cite{Poisson:1995sv} for the stability of the
Morris-Thorne WH.

We work with the tortoise coordinate $|dr/dr_{*}|=B$. Generally,
we define $r_*(r_0)=0$, and will have $r_*>0$ and $r_*<0$ for both
sides at the throat, respectively. To illustrate the GW waveforms, we scatter a test wave packet, which satisfies the Klein-Gordon equation in the pinching WH background $\Box\Phi=0$. We expand $\Phi$ as $
\Phi=\Sigma_{lm}{Y_{lm}(\theta,\phi)\over r}\Psi_{lm}(r)$, and get
the Regge-Wheeler Eq.
\begin{align}
\lf[-\frac{\partial^{2}}{\partial
t^{2}}+\frac{\partial^{2}}{\partial
r_{*}^{2}}-V_{l}(t,r_*)\rt]\Psi_{lm}(t,r_*)=0 ,
\label{Phi-equation}\end{align}
with
\ba V_l(t,r_*)= \left\{
                        \begin{array}{ll}  V_l^{BH}(r_*-L/2) \quad\, {\rm for} \quad r_*>0, \\
                                    V_l^{BH}(-r_*-L/2) \quad\, {\rm for} \quad r_*<0,
                \end{array}
                \right.
\ea where $V_l^{BH}(r_*(r))$ is the barrier $V_l^{BH}(r)=B
\lf[\frac{l(l+1)}{r^{2}}+\frac{B'}{r}\rt]$ of BH but written in
the coordinate $r_*^{BH}$. As an illustration, we will focus
on $l= 2$ in the following. In Schwarzschild-like WH
background, what $\Phi$ feels is a pair of mirror
potentials $V_l^{BH}(r_*(r))$ glued at $r_0$ ($r_*=0$), and the
separation between the barriers of mirror potentials is
\be L\simeq 2\int_{r_{0}}^{3M}\frac{dr}{B}\simeq
4M\log\lf[\frac{M} {\ell(t)}\rt], \quad\, {\rm for}\,\,
\ell(t)=r_{0}-2M\ll M, \label{L}\ee which will slowly get longer
for ${\dot \ell(t)}< 0$. When $\ell(t)=0$, $r_0$ equals to the
Schwarzschild radius, and the WH becomes a BH.


It has been found in Refs.\cite{Cardoso:2016rao, Cardoso:2016oxy}
that if the post-merger object is a WH, the ringdown
waveform will consist of the primary signal (almost identical to
that of BH) and a series of equal-interval echoes, see also
\cite{Nakano:2017fvh, Mark:2017dnq}. Considering the pinching of
WH is enough slow, we solve Eq.(\ref{Phi-equation}) with the
initial Gaussian perturbation
\begin{align}
\frac{\partial\Psi_{lm}}{\partial
t}(0,r)=e^{-(r_{*}-r_{g})^{2}/\sigma^{2}},\quad\,
\Psi_{lm}(0,r)=0,
\end{align}
where $r_{g}=10M$, $\sigma=6M$. We plot the corresponding
waveforms in Fig.2, and see that contrary to
Refs.\cite{Cardoso:2016rao, Cardoso:2016oxy}, the interval $\Delta
t_{echo}$ of the echoes in our scenario are not equal, but
increase with time.
The shift of interval following the $i^{th}$ echo $\delta t_i$
is approximately
\be \delta
t_{i}=\Delta t^{i+1,i}_{echo}-\Delta t^{i,i-1}_{echo}\sim
8M\log\lf[{\ell(t_i)\over \ell(t_{i+1})}\rt], \label{deltat}\ee
where $\Delta t^{i+1,i}_{echo}$ is the interval $\Delta t_{echo}$
between the $(i+1)^{th}$ and $i^{th}$ echoes.


We will estimate the quasinormal frequencies (QNFs)
in slowly pinching WH background. We focus on a period $t\simeq
2L$, during which the separation $L$ between the barriers will
become ${\tilde L} =L+\Delta L$. In the approximation
$\Delta L\ll L$, we could regard the moving of barriers as the
perturbation for the QNFs $\omega_{L}$, which will give
rise to the shifts of $\omega_{L}$ to $\omega_{{\tilde L}}$. Thus
in the frequency domain, we can write
Eq.(\ref{Phi-equation}) as \be \lf[\frac{\partial^{2}}{\partial
r_{*}^{2}}+{\omega}^2_{L}-V_{l}(L,r_*)\rt]{\hat \Psi}_{lm}=0,
\label{RW1}\ee where \be {\omega}^2_{L}=\omega^2_{\tilde
L}-{\partial \omega^2_L\over
\partial L}\Delta L, \label{omega}\ee and $\Psi(t,r_*)=\int
\frac{d\omega}{2\pi} {\hat{\Psi}}(\omega,r_*)e^{i\omega
t}$. In Eq.(\ref{RW1}), the separation between the barriers is
still $L$, the effect of $\Delta L$ ($\ll L$) is absorbed into
$\omega_{L}^2$. Eq.(\ref{RW1}) is the Regge-Wheeler equation for
the static WH, and its QNFs have been calculated in
Ref.\cite{Bueno:2017hyj}, \be \omega_{L,n}= {n\pi\over
L}+{i\ln\lf|R_{BH}(\omega_{L,n})\rt|\over L}, \ee
\begin{figure}[htbp]
\includegraphics[width=0.48\textwidth]{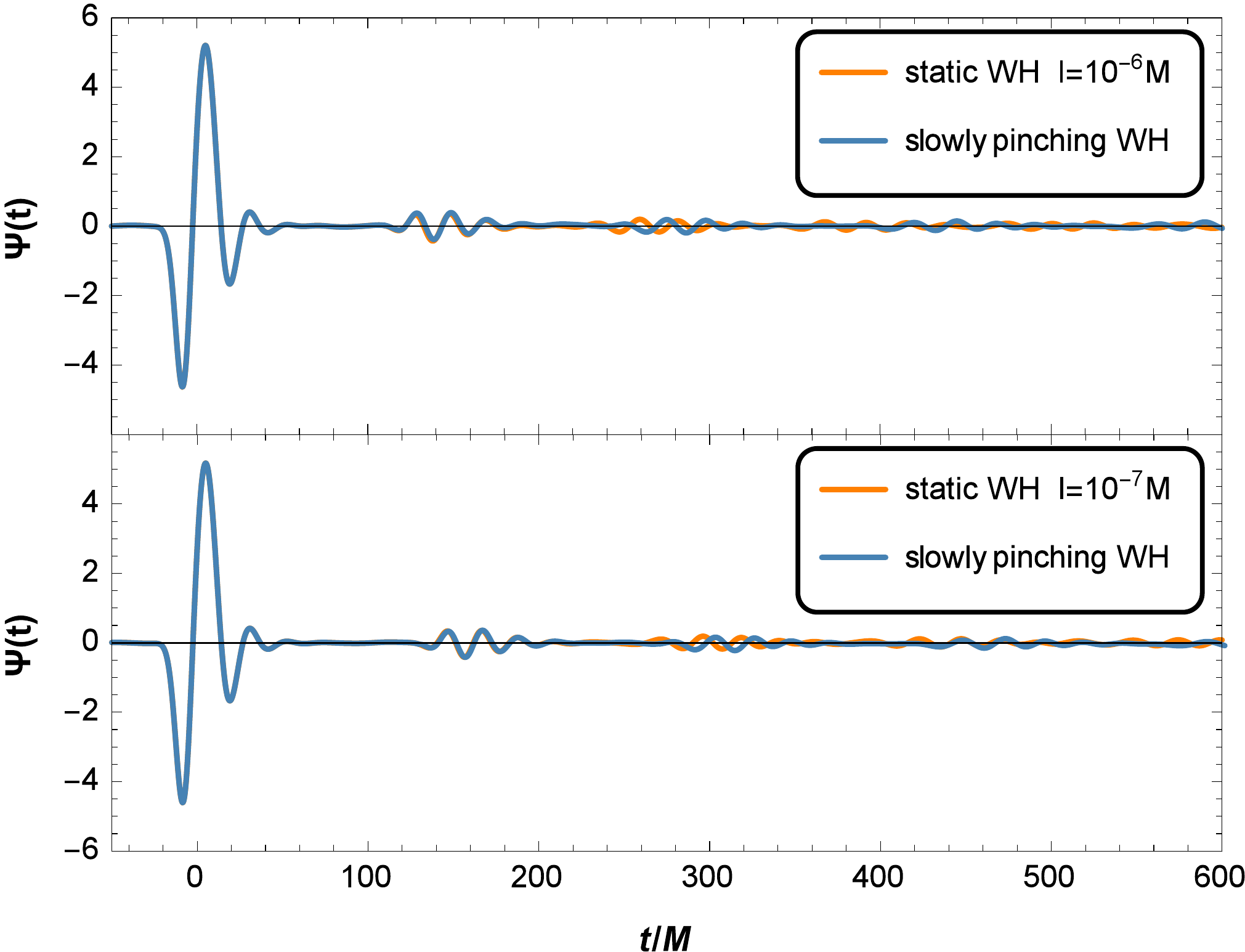}
\includegraphics[width=0.49\textwidth]{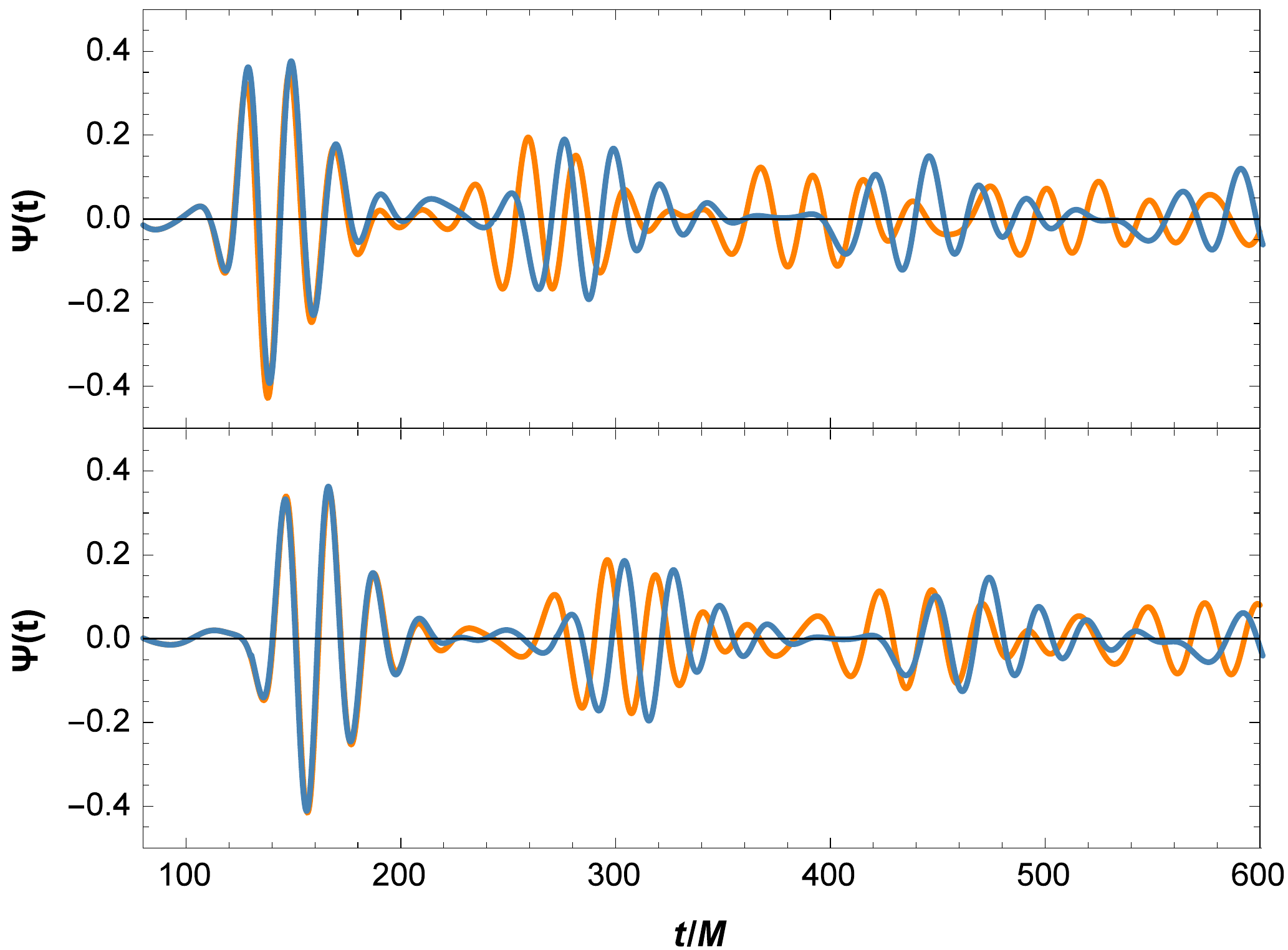}
\caption{Ringdown waveforms of post-merger objects, which
correspond to the static and slowly pinching WHs (depicted by
Fig.\ref{fig02}), respectively. The right panel is equivalent to
the left panel in the segment $100\leqslant t/M\leqslant 600$.}
\label{fig01}
\end{figure}
where $R_{BH}(\omega_{L,n})$ is the reflection coefficient of the
barrier $V_l^{BH}(r_*)$. Considering the expansion \be
R_{BH}(\omega_{L,n})=-1 +\sum_{j=1}^{\infty} {R^{(j)}_{BH}(0)\over
j!} \omega_n^k,\ee we have \be{Re}({\omega}_{L,n})\simeq
{ n\pi\over L},\quad\, {
Im}({\omega}_{L,n})={\ln\lf|R_{BH}(\omega_{L,n})\rt|\over
L}\lesssim {\cal O}({1\over L^3}). \ee According to
Eq.(\ref{omega}), we have \be \omega^2_{{\tilde L},n}\simeq
 \omega_{L,n}^2-2\omega^2_{{L},n}{\Delta L\over L}, \ee
where $\omega_{{L},n}\simeq n\pi/L$ is used.

Generally,
after the primary signal is reflected off the barrier on
the other side, the corresponding signal will consist of a
sum of WH QNMs, e.g.\cite{Bueno:2017hyj}. We find that in
slowly pinching WH background, after a period $t\simeq 2{L}$, the
QNFs reduce to \be {Re}({\omega}_{{\tilde L},n})\simeq {
n\pi\over L}\lf(1-{\Delta L\over
 L}\rt). \ee
Thus
\be \Psi(t)\sim \sum_{n=-\infty}^{\infty}c_{n}e^{-i({{
n}\pi\over L}-{{n}\pi\Delta L\over L^2})t} e^{{\rm
Im}(\omega_{L, n})t} =\sum_{n=-\infty}^{\infty}c_{
n}e^{-i2n\pi ({t\over \Delta t_{echo}})} e^{{\rm Im}(\omega_{L,
n})t} \label{Psi}\ee with the period \be \Delta t_{echo}\simeq
2L/\lf(1-{\Delta L\over
 L}\rt).\ee Thus
the signal will be repeated periodically ( referred to as
``echoes'' in the literature). However, since the WH is
slowly pinching off, we actually have $L_{i+1}=L_{i}+\Delta
L_{i}$ in successive period $t\simeq 2L_i$, so \be \Delta
t^{i+1,i}_{echo}\simeq \Delta t^{i,i-1}_{echo}/\lf(1-{\Delta
L_{i}\over L_i}\rt)>{\Delta
t^{i,i-1}_{echo}}.\label{DeltaL1}\ee Replacing $\Delta
t_{echo}$ in (\ref{Psi}) with $\Delta t^{i+1,i}_{echo}$, we will
obtain a waveform $\Psi(t)$ with the interval-increasing echoes.
Considering $\delta t_{i}=\Delta t^{i+1,i}_{echo}- \Delta
t^{i,i-1}_{echo}$ and $2L_i=\Delta t^{i,i-1}_{echo}$, we have
$\delta t_{i}\simeq 2\Delta L_{i}$, which is
consistent with Eqs.(\ref{L}) and (\ref{deltat}).


\section{Effect of the Interval Shift}

We will assess and discuss the effect of the shift of echo
interval on the search for the signals of echoes in GW data.
Based on Eqs.(\ref{Psi}) and (\ref{DeltaL1}),
the GWs ringdown waveform in Fig.\ref{fig01} is modelled as
\ba
\Psi(t) & =& \Psi^{BH}(t)+\Psi^{echo}(t)\nn\\
&=& {{\cal A}}e^{- t/{\tau}}\cos(2\pi {f}t+{\phi})+
\sum_{n=1}^{{\tilde N}_{echo}}(-1)^{n}{\cal
A}_{n}e^{-\frac{x^2_{n}}{2 \sigma^2_{n}}}\cos(2\pi {f}_n
x_n), \label{template} \ea where $\Psi^{BH}(t)$ is the post-merger
BH-like signal with the amplitude ${\cal A}$ and the damping time
$\tau$, and $\Psi^{echo}(t)$ is the echo signal with the
amplitude ${\cal A}_n\sim\frac{1}{3+n} {{\cal A}}$, which
is modulated by a Gaussian profile with the width
$\sigma_n$, and $x_n = t-\sum_{i=0}^n\Delta t^{i+1,i}_{echo}$.


When the signal and (\ref{template}) are maximally matched,
the expected matched-filter SNR is \cite{Allen:2005fk,
TheLIGOScientific:2016qqj}
\begin{equation}
\rho = \sqrt{4\int_{0}^{\infty}df\frac{|{\tilde
\Psi}(f)|^2}{S_{n}(f)}},
\end{equation}
where ${\tilde \Psi}(f)=\int \Psi(t)e^{-2\pi i ft}dt$, and
$S_{n}(f)$ is the noise power spectral density (PSD) of detector.
We focus on the GW1509014 event ($M\simeq
68M_{\bigodot}$), which yields ${f}\simeq 250$Hz and ${\tau}\simeq
4\times10^{-3}$s in (\ref{template})
\cite{TheLIGOScientific:2016src}. We also choose ${\cal
A}\simeq 6\times 10^{22}$, which is consistent with the best-fit
parameters for the GW150914 event. We, for simplicity, set all
$\sigma_n$ ($=\sigma$) as well as $\delta t_i$ ($=\delta
t$) equal,
and have \be x_n=t-(n+1)\Delta
t_{echo}^{1,0}-\frac{n(n+1)}{2}\delta t.\ee Regarding the
post-merger object of GW150914 as a pinching WH, we have $\Delta
t^{1,0}_{echo} \simeq 3\times 10^{-2}$s for initial $\ell(t)\sim
10^{-5}$.

\begin{figure}[htbp]
\includegraphics[scale=2,width=0.47\textwidth]{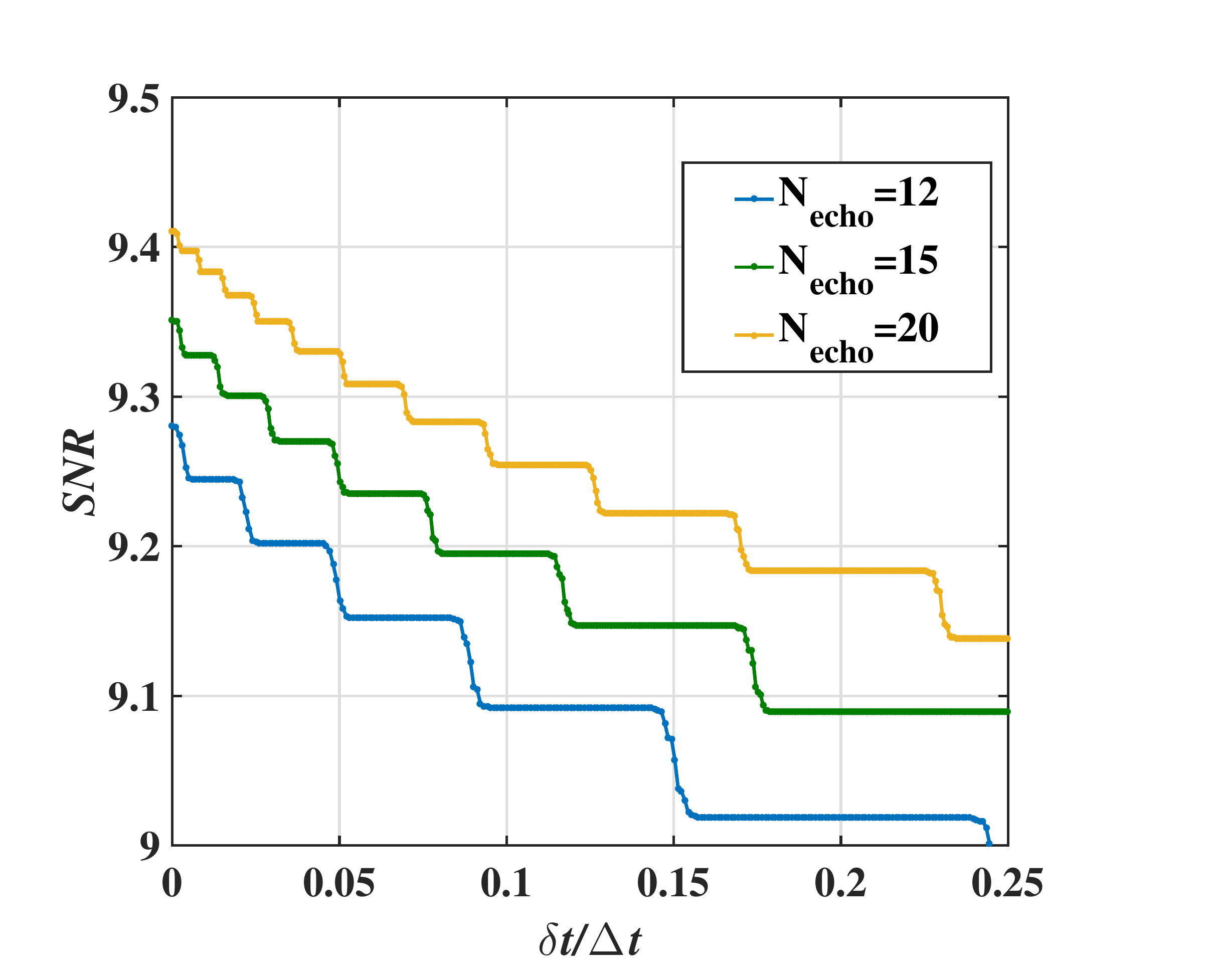}
\includegraphics[scale=2,width=0.52\textwidth]{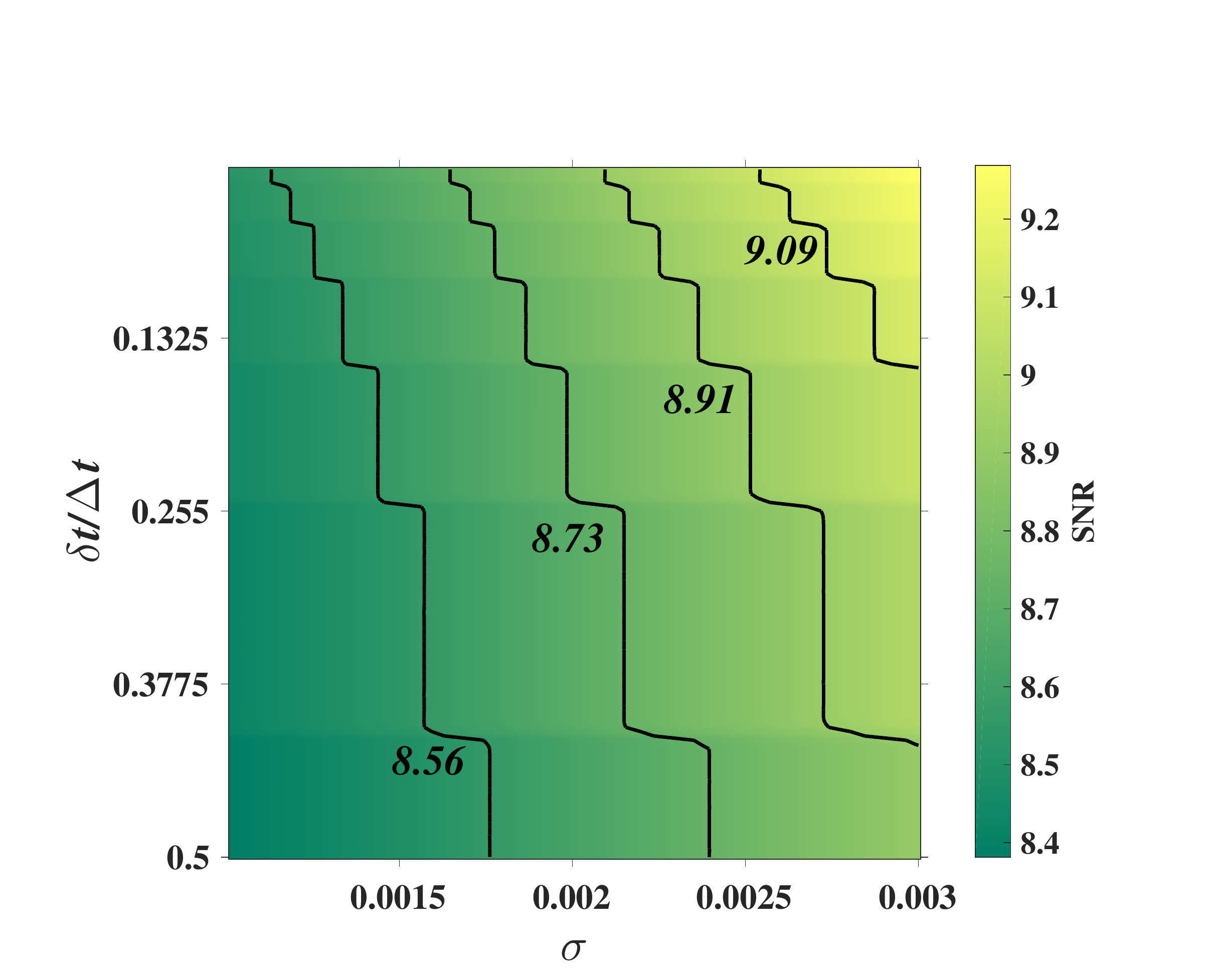}
\caption{Left panel: the SNR with respect to $\delta t/\Delta t$
for different $N_{echo}$. Right panel: the SNR with respect to
$\sigma$ and $\delta t/\Delta t$, we fix $N_{echo}=12$. }
\label{SNR}
\end{figure}

We calculate the SNR in a fixed segment $T=N_{echo}\Delta t\sim
3N_{echo}\times 10^{-2}$s. We plot the SNR with respect to $\delta
t/\Delta t$ in left panel of Fig.\ref{SNR}, where $\Delta t=\Delta
t_{echo}^{1,0}$ is set. We take $N_{echo}=20$ (so $T\simeq
3N_{echo}\times 10^{-2}$s$=0.6$s), and see that if all echoes are
equal-interval ($\delta t=0$), i.e.${\tilde N}_{echo}=N_{echo}$
for (\ref{template}), we have the SNR $\rho\simeq 9.4$, but if
$\delta t/\Delta t\simeq 0.1$, the SNR will reduce to $\rho\simeq
9.3$, since we only have ${\tilde N}_{echo}\simeq 12$ in this
segment. Thus the larger is $\delta t/\Delta t$, the less is the
number of echoes in fixed segment, so the lower is the SNR.
In right panel of Fig.\ref{SNR}, we also show how the
different values of $\delta t/\Delta t$ alter the SNR of the
signals with the echo width $\sigma$.

The shift of echo interval is encoded in $\delta t$.
 The LIGO/Virgo collaborations modelled the ringdown
waveform without the echoes as $\Psi^{BH}(t)$, see
(\ref{template}), and found the SNR $\rho\sim 8$
\cite{TheLIGOScientific:2016src}. Generally, the inclusion of
echoes will enhance the SNR, e.g.\cite{Maselli:2017tfq}. Our
result indicates that the shift of echo interval could
significantly affect the parameter estimation of echo signals,
when one searched for the corresponding signals in GW data.


\section{Discussion}


Even though after the merger a BH/BH binary (or BH/WH
binary) eventually develop into a BH, an exotic
intermediate state might exist. We show that if such a state is a
WH, which is slowly pinching off (and eventually will collapse
into a BH), the ringdown waveform will exhibit a series of
echoes, as pointed out in \cite{Cardoso:2016rao}. However,
we have found that the usual assumption that the GW echoes
are equal-interval is not always applicable.
In particular, in our scenario the intervals between the
neighboring echoes will increase with time. We have
argued the significant effect of the shift of echo interval on the
search for the signals of echoes in GW data released by
LIGO/Virgo.


The viability of WH depends on special models, which is
still a developing subject,
e.g.\cite{ArmendarizPicon:2002km,Rubakov:2015gza,Garcia:2011aa,Lobo:2016zle}.
Some of the issues might be better understood by performing
numerical simulations of binary mergers with WHs.
 The physics of GW echoes has recently been extensively studied, see also
\cite{Price:2017cjr,Zhang:2017jze,Conklin:2017lwb}. While
the post-merger object we considered is a WH, our result
 may also be applicable for other exotic compact objects
(e.g.
\cite{Mazur:2001fv,Visser:2003ge}\cite{Carballo-Rubio:2017tlh}),
as well as the BHs with the correction of modified/quantum gravity
\cite{Giddings:2017jts,Barcelo:2017lnx}, with the shift of their
reflector surface towards the Schwarzschild radius. However, if
initial state is not a BH, the inspiral stage could in principle
be used to discriminate against a two-BH initial state, since the
quadrupole moment, tidal love numbers or absorption of the initial
state is different from that of a BH, see
e.g.\cite{Cardoso:2017cqb}\cite{Maselli:2017cmm}.

\textbf{Acknowledgments}

We would like to thank Raul Carballo-Rubio and Leo C. Stein for
valuable comments, and Zhoujian Cao, Bin Hu for discussions. YSP
is supported by NSFC, Nos.11575188, 11690021, and also supported
by the Strategic Priority Research Program of CAS, No.XDB23010100.
YTW is supported in part by the sixty-second batch of China
Postdoctoral Fund. SYZ acknowledges support from the starting
grant of USTC and the 1000 Young Talent Program of China. JZ is
supported by the National Science and Engineering Research Council
through a Discovery grant.





 \end{document}